\documentclass{elsart}

\usepackage{epsfig}

\newcommand{\be}{\begin{equation}}
\newcommand{\ee}{\end{equation}}
\newcommand{\bea}{\begin{eqnarray}}
\newcommand{\eea}{\end{eqnarray}}

 \def\bean{\begin{eqnarray*}}
 \def\eean{\end{eqnarray*}}

 \def\gsim{\mathrel{\rlap{\lower0.2em\hbox{$\sim$}}\raise0.2em\hbox{$>$}}}
 \def\ksim{\mathrel{\rlap{\lower0.2em\hbox{$\sim$}}\raise0.2em\hbox{$<$}}}

\begin{document}

\begin{frontmatter}
\title{Strangeness production within Parton-Hadron-String Dynamics (PHSD)}
\author[unif]{O. Linnyk}, 
\author[unif]{E. L. Bratkovskaya},
\author[unig]{W. Cassing}
\address[unif]{Institut f\"ur Theoretische Physik, %
  Universit\"{a}t Frankfurt, 60438 Frankfurt am Main,
  Germany}
\address[unig]{Institut f\"ur Theoretische Physik, %
  Universit\"at Giessen, %
  35392 Giessen, %
  Germany}

\begin{abstract}
The Parton-Hadron-String Dynamics (PHSD) transport approach consistently
simulates the full evolution of a relativistic heavy-ion collision
from the initial hard scatterings  string formation through the
dynamical deconfinement phase transition to the quark gluon plasma (QGP),
to the hadronization and to  subsequent interactions in the
hadronic phase. The transport theoretical description of quarks and
gluons is based on a dynamical quasiparticle model for partons
matched to reproduce recent lattice QCD results in thermodynamic
equilibrium. The transition from partonic to hadronic degrees of
freedom is described by covariant transition rates for the fusion of
quark-antiquark pairs or three quarks (antiquarks). Studying Pb+Pb
reactions from 40 to 158 A$\cdot$GeV, we find that at most 40\% of
the collision energy is stored in the dynamics of the partons.
This implies that a large fraction of non-partonic, i.e. hadronic or
string-like matter, which can be viewed as a hadronic corona, is
present in these reactions, thus
neither hadronic nor purely partonic models can be employed to
extract physical conclusions in comparing model results with data.
On the other hand, comparing the PHSD results to those of the
Hadron-String Dynamics (HSD) approach without the phase transition
to QGP, we observe that the existence of the partonic phase has a
sizeable influence on the transverse mass distribution of final kaons
due to the repulsive partonic mean fields and initial partonic
scattering. Furthermore, we find a significant effect of the QGP on the
production of multi-strange antibaryons due to a slightly enhanced
$s\bar s$ pair production in the partonic phase from massive time-like
gluon decay and to a more abundant formation of strange antibaryons in
the hadronization process.
\end{abstract}

\end{frontmatter}
\vspace{-5pt} The nature of confinement and the phase transition
from a partonic system of quarks, antiquarks and gluons -- a
quark-gluon plasma (QGP) -- to interacting hadrons, as occurring in
relativistic nucleus-nucleus collisions, is a central topic of
modern high energy physics. In the present work, the dynamical
evolution of the heavy-ion collision is described by the PHSD
transport approach~\cite{CasBrat} incorporating the off-shell
propagation of the partonic quasi-particles according
to Ref.~\cite{Juchem} as well as the transition to resonant hadronic
states (or strings). Here we employ the PHSD approach -- described
in~\cite{CasBrat,Cassing:2009vt} -- to strangeness production in
nucleus-nucleus collisions at moderate relativistic energies, i.e.
at SPS energies up to 160~A$\cdot$GeV.

\vspace{-5pt} {\bf The PHSD approach} \vspace{-5pt}

A consistent dynamical approach -- valid also for strongly
interacting systems -- can be formulated on the basis of the
Kadanoff-Baym equations \cite{KBaym,Sascha1} or off-shell transport
equations in phase-space representation, respectively
\cite{Juchem,Sascha1}. In the Kadanoff-Baym theory the field quanta
are described in terms of propagators with complex selfenergies.
Whereas the real part of the selfenergies can be related to
mean-field potentials, the imaginary parts  provide information
about the lifetime and/or reaction rates of time-like 'particles'
\cite{Andre}. Once the proper (complex) selfenergies of the degrees
of freedom are known, the time evolution of the system is fully
governed  by off-shell transport equations (as described in Refs.
\cite{Juchem,Sascha1}).

The PHSD approach is a microscopic covariant transport model that
incorporates effective partonic as well as hadronic degrees of
freedom and involves a dynamical description of the hadroni\-zation
process from partonic to hadronic matter \cite{CasBrat}. Whereas the
hadronic part is essentially equivalent to the conventional HSD
approach \cite{HSD} the partonic dynamics is based on the Dynamical
QuasiParticle Model (DQPM) \cite{Cassing06} which describes QCD
properties in terms of single-particle Green's functions (in the
sense of a two-particle irreducible approach) and leads to effective
strongly interacting partonic quasiparticles with broad spectral
functions as degrees of freedom.

The off-shell parton dynamics also allows for a solution of the
hadronization problem: the hadronization occurs by quark-antiquark
fusion or 3 quark/3 antiquark recombination which is described by
covariant transition rates as introduced in
Ref.~\cite{CasBrat,Cassing:2009vt}, obeying flavor
current-conservation, color neutrality as well as energy-momentum
conservation. Since the dynamical quarks become very massive close
to $T_c$, the formed resonant 'pre-hadronic' color-dipole states
($q\bar{q}$ or $qqq$) are of high invariant mass, too, and
sequentially decay to the ground state meson and baryon octets
increasing the total entropy. This solves the entropy problem in
hadronization in a natural way~\cite{CassXing}.
\begin{figure}
  \begin{minipage}[b]{0.45\textwidth}
    \includegraphics*[width=\textwidth]{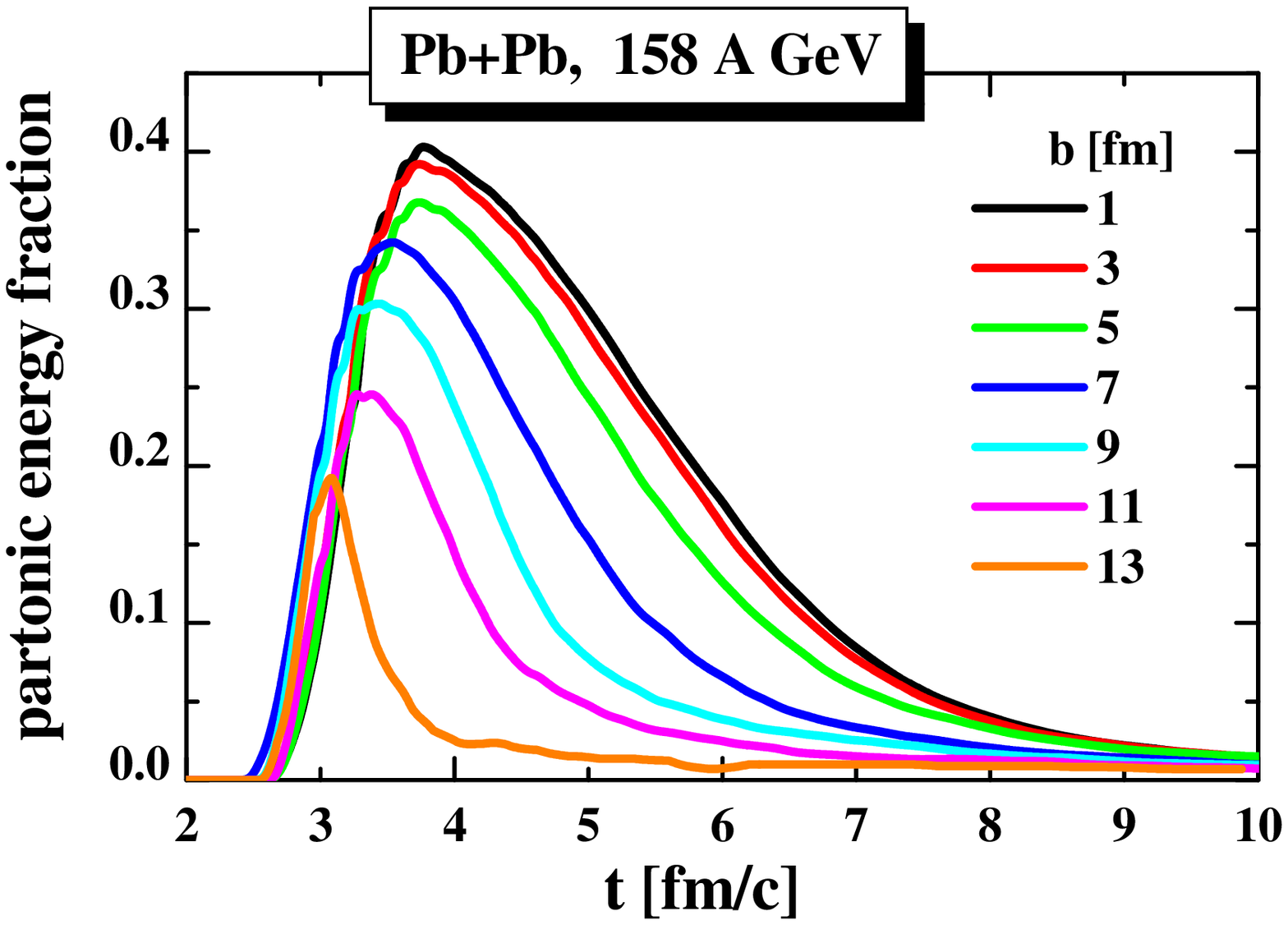} \caption{The partonic energy fraction
 as a function of time for impact
parameters $b$ from 1 fm to 13 fm in steps of 2 fm for Pb+Pb at 158
A$\cdot$GeV.}  \label{fig11}
  \end{minipage}
  \hspace{0.09\textwidth}
  \begin{minipage}[b]{0.45\textwidth}
    \includegraphics*[width=\textwidth]{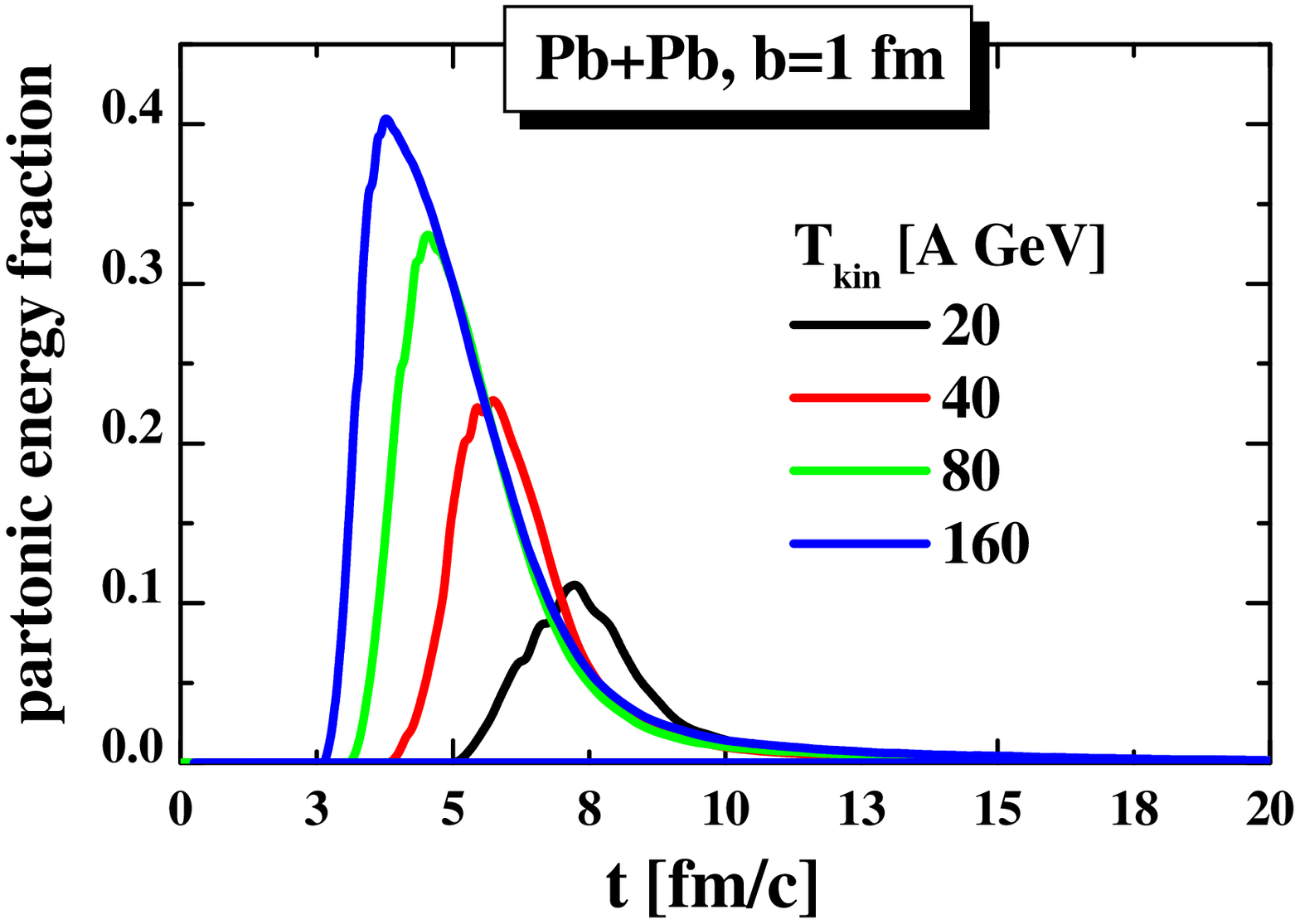}
\caption{The partonic energy fraction as a function of time for
impact parameter $b$ = 1 fm for Pb+Pb at 160, 80, 40 and
20~A$\cdot$GeV.}  \label{fig10}
  \end{minipage}
\end{figure}

\vspace{-5pt} {\bf Partonic energy fractions} \vspace{-5pt}

We start with a consideration of energy partitions in order to map
out the fraction of partonic energy in time for relativistic
nucleus-nucleus collisions. The question emerges to what extent the
partonic phase shows up as a function of centrality. To this aim we
study the partonic energy $E_p(b)$ relative to the energy converted
from the kinetic motion of the impinging nuclei at impact parameter
$b$, i.e. $ 
R_p (b) = E_p(b)/(E_B(t=0,b)-E_B(t\rightarrow \infty,b)) , $ which
is displayed in Fig.~\ref{fig11} for impact parameter $b$ from 1~fm
to 13~fm in steps of 2~fm. The system is Pb+Pb at 158 A$\cdot$GeV.
One observes that even at very central collisions the partonic
energy fraction $R_p$ only reaches about 40\% and decreases to about
20\% (in the peak) for very peripheral reactions. More striking is
the fact that the duration of the partonic phase shrinks
substantially when going from central to peripheral reactions. Thus
the 'popular picture' that a partonic phase is reached in the
overlap region of the nuclei at top SPS energies is by far not
substantiated by the PHSD calculations. This has been been pointed
out before in Refs.~\cite{Werner
}. In conclusion, very
central collisions show a sizeable fraction of space-time regions of
partonic nature, but also of the hadronic (or string-like) corona.
At first sight this looks discouraging, but one should concentrate
on observables with a special sensitivity to the partonic phase,
e.g. strangeness~\cite{Cassing:2009vt},
dileptons~\cite{Linnyk:2009nx}, charm~\cite{Linnyk:2008hp} or
jets~\cite{Gallmeister:2004iz}.

We continue with the ratio $R_p$ as a function of bombarding energy
concentrating here on the FAIR and full SPS energy regime but
considering only central collisions ($b$ = 1 fm). The results are
displayed in Fig. \ref{fig10} for central Pb+Pb collisions at 160,
80, 40, and 20~A$\cdot$GeV as a function of time and demonstrate
that the average duration of the partonic phase does not change very
much with bombarding energy, however, the partonic volume shrinks by
about a factor of three when stepping down in bombarding energy from
160 to 20~A$\cdot$GeV. Thus according to our PHSD calculations there
should be a QGP also at FAIR energies but its space-time volume is
significantly smaller than that for the hadronic phase.

{\bf Particle spectra in comparison to experiment} \vspace{-5pt}
\begin{figure}
{\psfig{figure=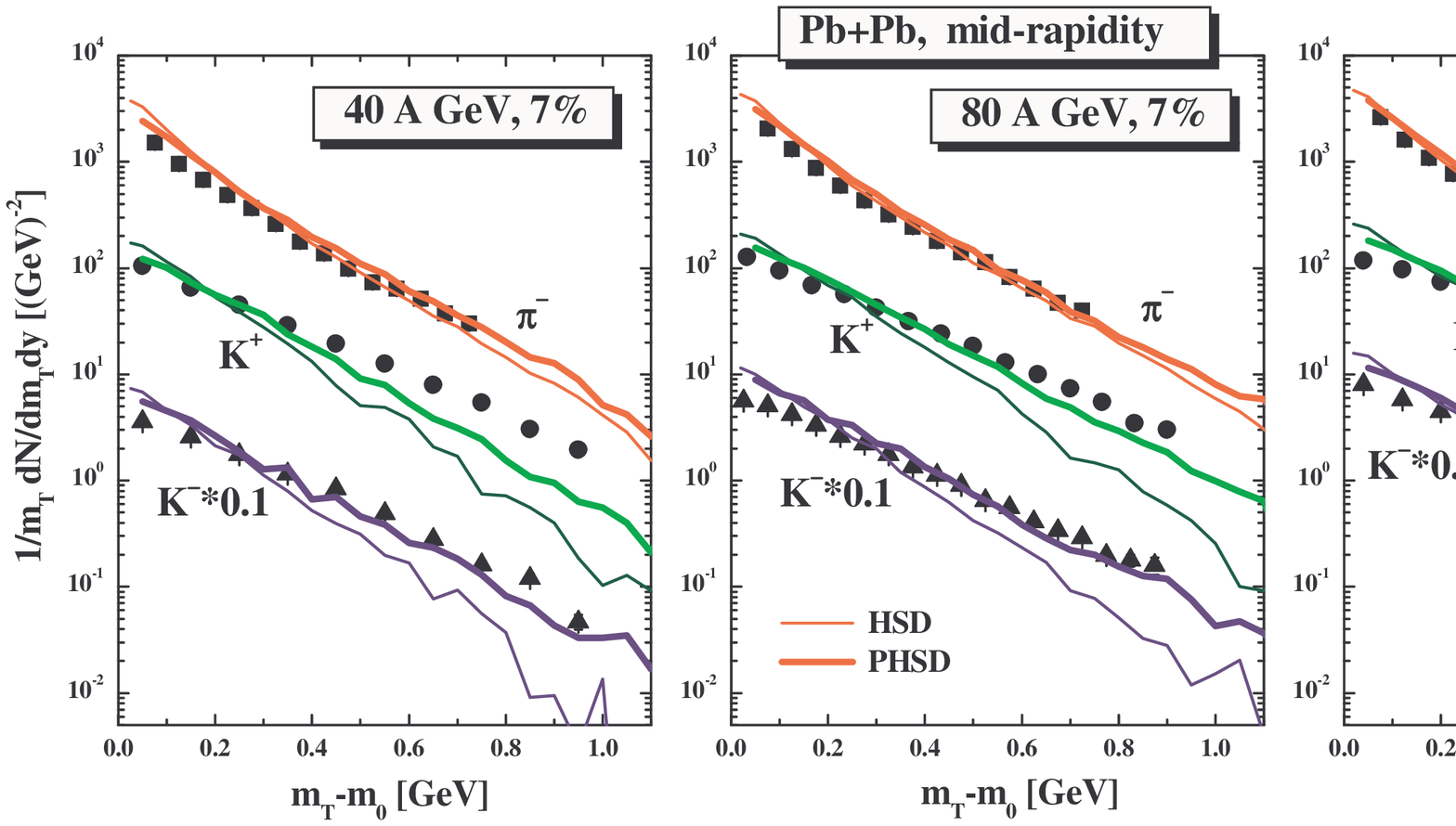,width=0.7\textwidth}} \caption{The
$\pi^-$, $K^+$ and $K^-$ transverse mass spectra for central Pb+Pb
collisions at 40, 80 and 158 A$\cdot$GeV from PHSD (thick solid
lines) in comparison to the distributions from HSD (thin solid
lines) and the experimental data from the NA49 Collaboration
\cite{NA49a}.  } \label{fig14}
\end{figure}

It is of interest, how the PHSD approach compares to the
HSD~\cite{HSD} model (without explicit partonic degrees of freedom)
as well as to experimental data. In Fig.~\ref{fig14} we show the
transverse mass spectra of $\pi^-$, $K^+$  and $K^-$ mesons for 7\%
central Pb+Pb collisions at 40 and 80 A$\cdot$GeV and 5\% central
collisions at 158 A$\cdot$GeV in comparison to the data of the NA49
Collaboration \cite{NA49a}.  Here the slope of the $\pi^-$ spectra
is only slightly enhanced in PHSD relative to HSD which demonstrates
that the pion transverse motion shows no sizeable sensitivity to the
partonic phase. However, the $K^\pm$ transverse mass spectra are
substantially hardened with respect to the HSD calculations at all
bombarding energies - i.e. PHSD is more in line with the data - and
thus suggest that partonic effects are better visible in the
strangeness-degrees of freedom. The hardening of the kaon spectra
can be traced back to initial parton-parton scattering as well as a
larger collective acceleration of the partons in the transverse
direction due to the presence of repulsive vector fields for the
partons.  We
recall that in Refs. \cite{BratPRL
} the
underestimation of the $K^\pm$ slope by HSD (and also UrQMD) had
been suggested to be a signature for missing partonic degrees of
freedom; our present PHSD calculations support this early
suggestion.
\begin{figure}
\centerline{\psfig{figure=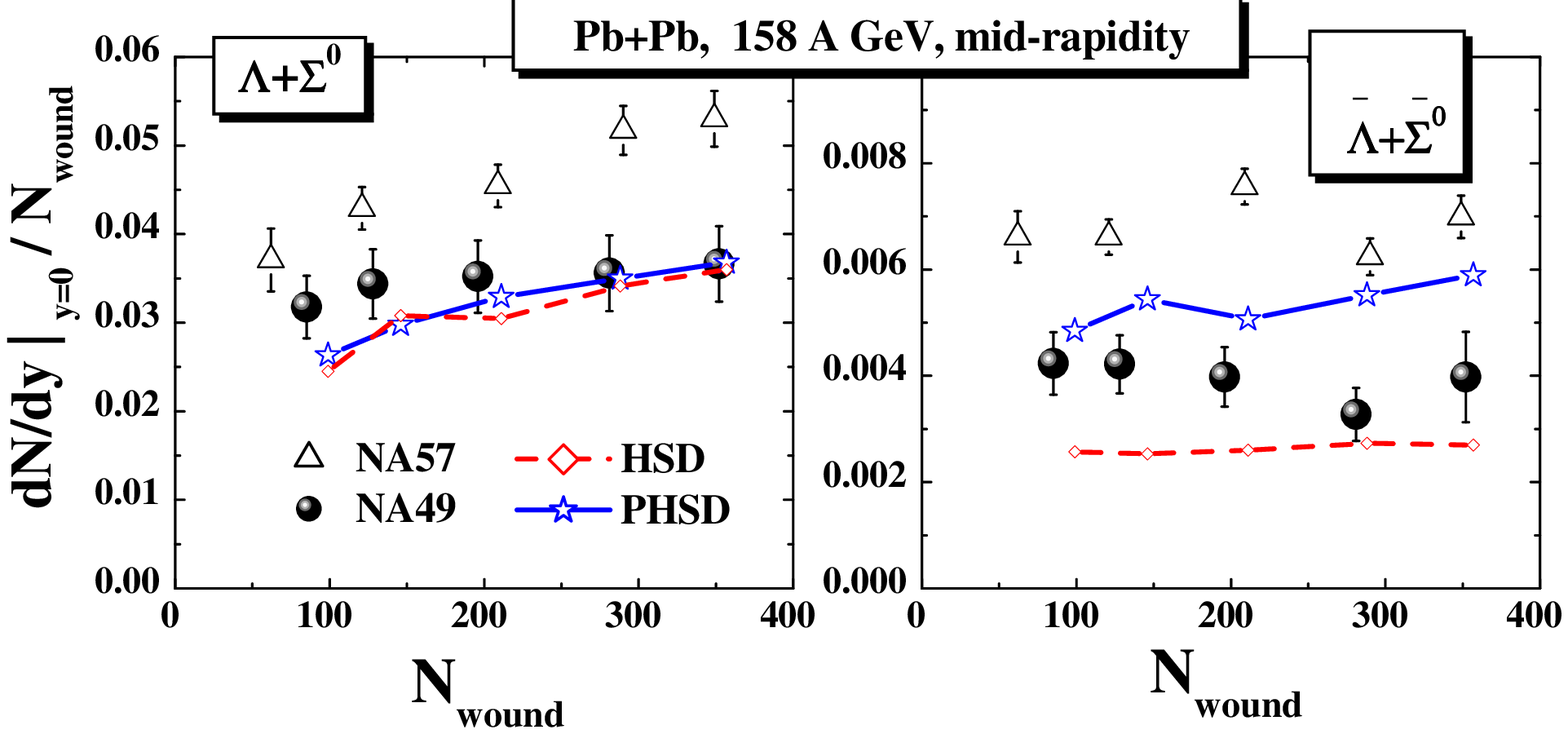,width=0.75\textwidth}}
\centerline{\psfig{figure=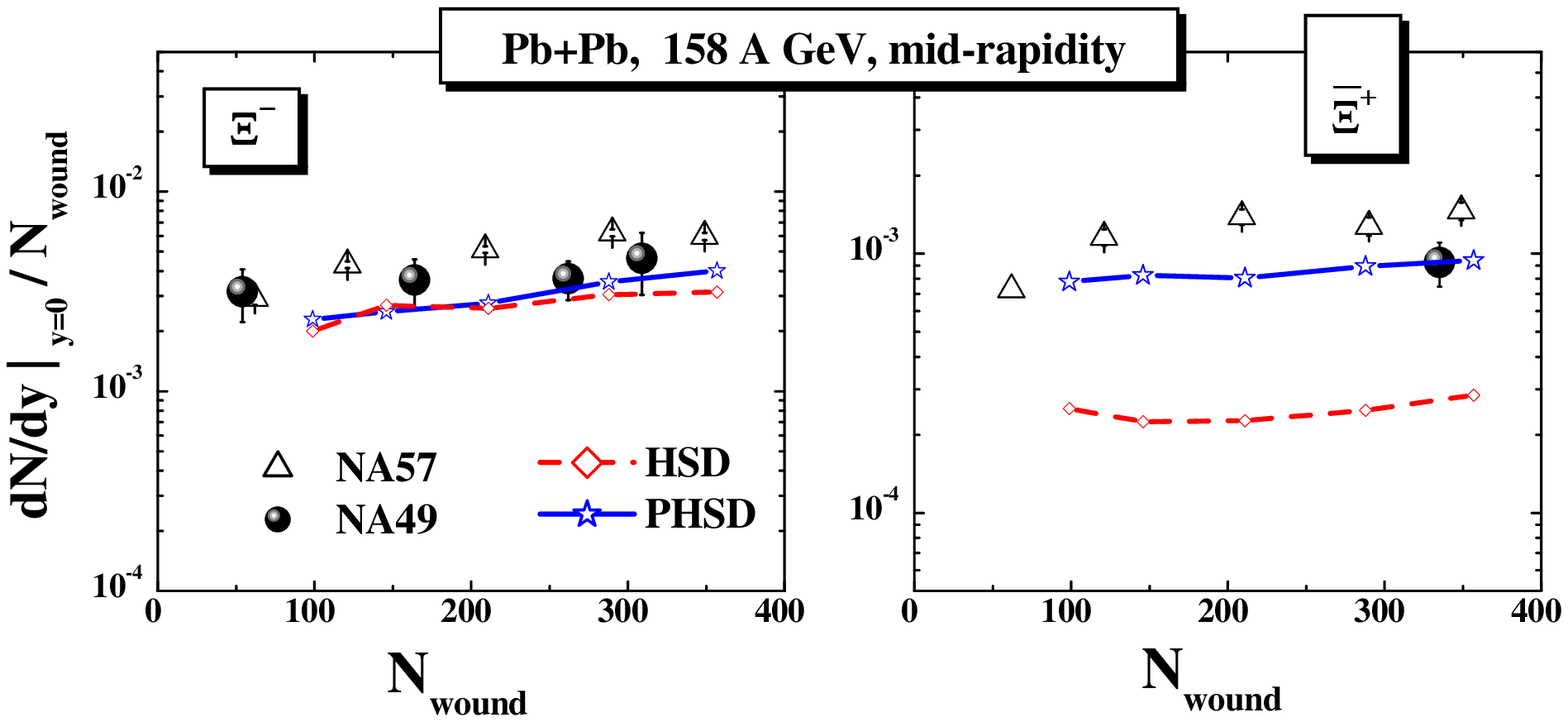,width=0.75\textwidth}}
\caption{{\bf Top:} The multiplicities of $(\Lambda +
\Sigma^0)/N_{wound}$ (l.h.s.) and $(\bar \Lambda + \bar
\Sigma^0)/N_{wound}$ (r.h.s.) as a function of the number of wounded
nucleons for Pb+Pb collisions at 158 A$\cdot$GeV at mid-rapidity
from PHSD (blue solid lines) and HSD (red dashed-dotted lines) in
comparison to the experimental data from the NA57 Collaboration
\cite{NA57} (open triangles) and the NA49 Collaboration
\cite{NA49_aL09} (solid dots). The calculations have an error of
5$-$10\% due to limited statistics. {\bf Bottom:} The multiplicities
of $\Xi^-/N_{wound}$ (l.h.s.) and $\bar \Xi^+/N_{wound}$ (r.h.s.) vs
$N_{wound}$ for Pb+Pb collisions at 158 A$\cdot$GeV at mid-rapidity.
Line coding as above.} \label{fig15d} \label{fig15e}
\end{figure}

The strange antibaryon sector is of further interest since here the
HSD calculations have always underestimated the yield \cite{Geiss}.
In this respect we compare in Fig.~\ref{fig15d} (top part) the
multiplicities of $(\Lambda + \Sigma^0)/N_{wound}$ (l.h.s.) and
$(\bar \Lambda + \bar \Sigma^0)/N_{wound}$ (r.h.s.) as functions of
the number of wounded nucleons $N_{wound}$ for Pb+Pb collisions at
158~A$\cdot$GeV at mid-rapidity from PHSD and HSD to the
experimental data from the NA57 Collaboration \cite{NA57} and the
NA49 Collaboration \cite{NA49_aL09}. Whereas the HSD and PHSD
calculations both give a reasonable description of the $\Lambda +
\Sigma^0$ yield of the NA49 Collaboration, both models underestimate
the NA57 data (open triangles) by about 30\%. An even larger
discrepancy in the data from the NA49 and NA57 Collaborations is
seen for $(\bar \Lambda + \bar \Sigma^0)/N_{wound}$ (r.h.s.); here
the PHSD calculations give results which are in between the NA49
data (solid dots) and the NA57 data (open triangles). We see that
HSD underestimates the $(\bar \Lambda + \bar \Sigma^0)$ midrapidity
yield at all centralities. This observation points towards a
partonic origin but needs further examination.

The latter result suggests that the partonic phase does not show up
explicitly in an enhanced production of strangeness (or in
particular strange mesons and baryons) but leads to a different
redistribution of antistrange quarks between mesons and antibaryons.
To examine this issue we show in Fig.~\ref{fig15e} (bottom part) the
multiplicities of  $\Xi^-$ baryons (l.h.s.) and $\bar \Xi^+$
antibaryons (r.h.s.) vs. $N_{wound}$ for Pb+Pb collisions at 158
A$\cdot$GeV at mid-rapidity from PHSD and HSD in comparison to the
experimental data from the NA57 Collaboration \cite{NA57} and the
NA49 Collaboration \cite{NA49b,NA49_aL09}. The situation is very
similar to the case of the strange baryons and antibaryons before:
we find no sizeable differences in the double strange baryons from
HSD and PHSD -- in a good agreement with the NA49 data -- but
observe a large enhancement in the double strange antibaryons for
PHSD relative to HSD.

\vspace{-5pt} {\bf Summary} \vspace{-5pt}

The PHSD approach has been applied to nucleus-nucleus collisions
from 20 to 160~A$\cdot$GeV in order to explore the space-time
regions of `partonic matter'. We have found that even central
collisions at the top SPS energy of $\sim$160~A$\cdot$GeV show a
large fraction of non-partonic matter. It is also found that though
the partonic phase has only a very low impact on rapidity
distributions of hadrons~\cite{Cassing:2009vt}, it has a sizeable
influence on the transverse-mass distribution of final kaons due to
the repulsive partonic mean fields and initial parton interactions. On the
other hand, the most pronounced effect of the partonic phase is seen
on the production of multi-strange antibaryons due to a slightly
enhanced $s{\bar s}$ pair production in the partonic phase from
massive time-like gluon decay and a more abundant formation of
strange antibaryons in the hadronization process. We also mention
that partonic production channels for dileptons appear to be visible
in the $\mu^+ \mu^-$ spectra from In+In collisions at
158~A$\cdot$GeV in the intermediate invariant mass
range~\cite{Linnyk:2009nx}.

Supported in part by the ``HIC for FAIR" center of the ``LOEWE''
program. \vspace{-20pt}
%
%


\begin{thebibliography}{99}
\vspace{-15pt}
\bibitem{CasBrat} W. Cassing and E. L. Bratkovskaya,
    {Phys. Rev.} C 78 (2008) 034919.
\bibitem{Juchem}
      W. Cassing and S. Juchem, {Nucl. Phys.} A 665 (2000) 377;
      {\it ibid} A 672 (2000) 417.
\bibitem{Cassing:2009vt}
  W.~Cassing and E.~L.~Bratkovskaya,
  Nucl.\ Phys.\  A {\bf 831} (2009) 215.
\bibitem{KBaym}
      L. P. Kadanoff, G. Baym, {\it Quantum Statistical Mechanics},
      Benjamin, 1962.
\bibitem{Sascha1}
      W. Cassing and S. Juchem, Nucl. Phys. A 672 (2000) 417;
    S. Juchem {\it et al.}, Nucl. Phys. A {743} {2004} 92.
\bibitem{Andre}
      A. Peshier and W. Cassing, { Phys. Rev. Lett.} 94 (2005) 172301.
\bibitem{HSD}
      W. Cassing and E. L. Bratkovskaya, { Phys. Rept.} 308 (1999) 65;
      W. Ehehalt and W. Cassing, {Nucl. Phys.} A 602 (1996) 449.
\bibitem{Cassing06}
      W. Cassing, {Nucl. Phys.} A 791 (2007) 365; {\it ibid.}  A 795 (2007) 70.
\bibitem{PRC08} W. Cassing and E. L. Bratkovskaya, Phys. Rev. C 78
(2008) 034919.
\bibitem{CassXing}
 W.~Cassing, E.~L.~Bratkovskaya and Y.~Xing, Prog. Part. Nucl.
 Phys. 62 (2009)  359.
\bibitem{Werner}
J.~Aichelin and K.~Werner,
  Phys.\ Rev.\  C {\bf 79}, 064907 (2009);
K. Werner, Phys. Rev. Lett. 98 (2007) 152301;
F. Beccattini and J. Manninen, J. Phys.
G 35 (2008) 104013.
\bibitem{Linnyk:2009nx}
  O.~Linnyk, E.~L.~Bratkovskaya and W.~Cassing,
  Nucl.\ Phys.\  A {\bf 830} (2009) 491C.
\bibitem{Linnyk:2008hp}
  O.~Linnyk, E.~L.~Bratkovskaya and W.~Cassing,
  Int.\ J.\ Mod.\ Phys.\  E {\bf 17} (2008) 1367.
\bibitem{Gallmeister:2004iz}
  K.~Gallmeister and W.~Cassing,
  Nucl.\ Phys.\  A {\bf 748} (2005) 241.
\bibitem{NA49a}
C. Alt et al., NA49 Collaboration, Phys. Rev. C 66 (2002) 054902;
Phys. Rev. C 77 (2008) 024903.
\bibitem{BratPRL}  E. L. Bratkovskaya, S. Soff, H. St\"ocker, M. van Leeuwen,
and W. Cassing, Phys. Rev. Lett. 92 (2004) 032302;
       E. L. Bratkovskaya, W. Cassing, and H. St\"ocker,
            Phys. Rev. C 67 (2003) 054905;
       E. L. Bratkovskaya {\it et al.},
            Phys. Rev.  { C 69} (2004) 054907.
\bibitem{Geiss}  J. Geiss, W. Cassing and C. Greiner, Nucl.
Phys. A 644 (1998) 107.
\bibitem{NA57} F. Antinori {\it et al.}, Phys. Lett. B 595 (2004) 68;
  J. Phys. G: Nucl. Phys. 32 (2006) 427.
\bibitem{NA49_aL09}
  T.~Anticic {\it et al.}  [NA49 Collaboration],
  Phys.\ Rev.\  C {\bf 80} (2009) 034906.
\bibitem{NA49b}
C. Alt et al., NA49 Collaboration, Phys. Rev. C 78 (2008) 034918.
\bibitem{Johann}  J. Rafelski and B. M\"uller, Phys. Rev.
Lett. 48 (1982) 1066.

\end{thebibliography}
\end{document}